\documentclass[anz]{cupaus}

\usepackage{graphicx}
\usepackage{amsmath}
\usepackage{amssymb}
\usepackage{bm}
\usepackage{bbold}
\usepackage{array}
\usepackage{colordvi}
\usepackage[numbers,sort&compress]{natbib}

\graphicspath{{.}{./EPS/}}

\newcolumntype {s}[1]{@{\hspace{#1}}} 
\newcolumntype {R}{>{$}r<{$}}         
\newcolumntype {C}{>{$}c<{$}}         
\newcolumntype {L}{>{$}l<{$}}         

\newcommand{\ket}[1]{\left | \, #1 \right \rangle}

\newcommand* {\openone}{\ensuremath{\mathbb{1}}}
\newcommand* {\ee}{\ensuremath{\mathrm{e}}}

\newcommand*{\vek}[1]{{\ensuremath{\bm{\mathrm{#1}}}}}

\newcommand*{\rr}{{\ensuremath{\bm{\mathrm{r}}}}}
\newcommand*{\pp}{{\ensuremath{\bm{\mathrm{p}}}}}
\newcommand* {\tvek}[2][c]{\left( \begin{array}{s{0.15em}#1s{0.15em}}
     #2\end{array} \right)}
\newcommand{\ir}{\mathcal{D}}

\definecolor{grey}{rgb}{0.4,0.4,0.4}

\begin{document}

\verso{}
\recto{}

\title{Discrete symmetries of low-dimensional Dirac models: A
 selective review with a focus on condensed-matter realisations}

\author[1]{R. Winkler}

\cauthormark 
\author[2]{U. Z{\"u}licke}

\address[1]{Department of Physics, Northern Illinois University, DeKalb,
Illinois 60115, USA and Institut f\"ur Theoretische Physik, Universit\"at
 Regensburg, 93040 Regensburg, Germany\email[1]{rwinkler@niu.edu}}

\address[2]{School of Chemical and Physical Sciences and
MacDiarmid Institute for Advanced Materials and Nanotechnology,
Victoria University of Wellington, PO Box 600, Wellington 6140,
New Zealand\email[2]{uli.zuelicke@vuw.ac.nz}}
      
\pages{1}{15}
      
\begin{abstract}
  The most fundamental characteristics of a physical system can
  often be deduced from its behaviour under discrete symmetry
  transformations such as time reversal, parity and chirality. Here
  we review basic symmetry properties of the relativistic quantum
  theories for free electrons in (2+1)- and (1+1)-dimensional spacetime.
  Additional flavour degrees of freedom are necessary to properly
  define symmetry operations in (2+1) dimensions and are generally
  present in physical realisations of such systems, e.g., in single
  sheets of graphite. We find that there exist two possibilities for
  defining any flavour-coupling discrete symmetry operation of the
  two-flavour (2+1)-dimensional Dirac theory. Physical implications
  of this previously unnoticed duplicity are discussed.
\end{abstract}

\keywords[2010 \textit{Mathematics subject classification}]{81R05}

\keywords[\textit{Keywords and phrases}]{relativistic quantum physics, reduced
dimensionality, Dirac Hamiltonians, quasi-relativity in solids}

\maketitle

\section{Introduction}\label{sec1}  
The opportunity to study aspects of two-dimensional (2D) relativistic
electron behaviour in condensed-matter systems~\cite{cas09, par09,
gib09, sin11, qi11, gom12}, using ultra-cold atoms~\cite{zhu07, wun08,
gol09, cir10, tar12}, and even in photonic structures~\cite{hal08, sep08}
has renewed interest in 2D versions of the Dirac equation~\cite{dir28, tha92}
for a free particle,
\begin{subequations}
  \label{eq:dirac}
  \begin{equation}
    \label{eq:dirac_eq}
    H_\mathrm{D} \psi = i\partial_t \psi,
  \end{equation}
  with Hamiltonian
  \begin{equation}\label{eq:dirac_ham}
    H_\mathrm{D} (\pp) = \vek{\alpha}\cdot\pp + \beta \, m \, .
  \end{equation}
\end{subequations}
We summarise here some features arising due to the reduced
dimensionality, focusing especially on symmetry properties.
Comparison is made between the previously considered
(2+1)-dimensional quantum electrodynamics
(QED$_3$)~\cite{jac81, app86, dit00} and theoretical studies of
quasi-relativistic condensed-matter systems~\cite{sem84, gus07,
cha08, sch08, kit09, ryu10, ber12, kos12}. We use units where the
Planck constant $\hbar=1$ and also the speed of light in vacuum
$c = 1$.

\subsection{Dirac Equation: Basics}
We employ the general expression for the Dirac Hamiltonian given in
Eq.~(\ref{eq:dirac_ham}) that is valid in any spatial dimensions $d \ge 1$.
Here $\vek{p}$ denotes the operator of the Dirac particle's momentum
which, in position ($\rr$) representation is given by $\vek{p}\equiv
-i \vek{\nabla}_\rr$. $\beta$ is a matrix, and $\vek{\alpha}$ is a vector of
matrices $\alpha_j$ that satisfy the relations~\cite{tha92, gre00}
\begin{subequations}\label{eq:matrels}
\begin{eqnarray}
\alpha_j\, \alpha_k + \alpha_k\, \alpha_j &=& 2\,\delta_{jk}\, \openone
\quad , \\
\alpha_j \, \beta + \beta\, \alpha_j &=& 0 \quad ,
\end{eqnarray}
with $\delta_{ik}$ denoting the familiar Kronecker symbol and $\openone$
being the $d$-dimensional identity matrix. Furthermore,
\begin{equation}
\beta^2 \equiv \alpha_j^2 = \openone \quad .
\end{equation}
\end{subequations}
While the explicit form of these matrices depends on the dimension, the
structure of the Hamiltonian $H_\mathrm{D}$ does not. For our future
discussion of symmetries, it will be useful to note that $\vek{\alpha}$ has
the physical interpretation of being the operator for the velocity $\vek{v}$
of a free Dirac particle. This can be seen from the basic relation defining
the time derivative of the position operator, $\vek{v} \equiv\dot{\rr} = i \,
[ H_\mathrm{D}\, , \vek{r}]$, for which application of the canonical
commutation relations $\, [ r_j \, , p_k ] = i \, \delta_{jk}$ straightforwardly
yields $\vek{v}\equiv \vek{\alpha}$.

The relativistically covariant form of the Dirac equation is
obtained by multiplying Eq.\ (\ref{eq:dirac_eq}) by
$\beta$ and re-arranging to exhibit derivatives w.r.t.\ time and
spatial coordinates in a unified fashion. Using the definitions
$\gamma^0 = \beta$, $\gamma^j = \beta \, v_j$ with $j=1,\dots,d$,
and $P_0=i\partial_{t}$, $P_j = -p_j\equiv i \partial_{r_j}$
for $j=1,\dots,d$ as the components of the covariant $(d+1)$-vector
of momentum~\cite{gre00}, one finds
\begin{equation}
\label{eq:dirac_op}
\left( \gamma^\mu \, P_\mu - m \right) \psi = 0 \quad ,
\end{equation}
where we have used the convention that repeated indices are summed
over. Application of the relations (\ref{eq:matrels}) yields the identities
\begin{subequations}
\begin{eqnarray}
\gamma^\mu\, \gamma^\nu + \gamma^\nu\, \gamma^\mu &=& 0
\quad \mbox{for $\mu\ne\nu$}\quad , \\
\left( \gamma^0 \right)^2 = - \left( \gamma^j \right)^2 &=& \openone \quad .
\end{eqnarray}
\end{subequations}

As the Dirac Hamiltonian for a free particle commutes with momentum
$\vek{p}$, we can choose plane waves as basis in real space and represent
$\vek{p}$ as an ordinary vector in $\mathbb{R}^d$. In the following,
we adhere to this convention of condensed-matter physics. The spectrum
of the Dirac Hamiltonian (\ref{eq:dirac_ham}) has two branches, one at
positive and one at negative energies. These are given by
\begin{equation}
E_\pm(\vek{p}) = \pm\sqrt{|\vek{p}|^2 + m^2}
\end{equation}
independent of the number $d$ of spatial dimensions. The eigenstates
in position representation have the general form
\begin{equation}\label{eq:eigenstates}
\psi_{\vek{p}}^\pm(\vek{r}) = \ee^{i\vek{p}\cdot\vek{r}}\,
\chi_{\vek{p}}^\pm \quad .
\end{equation}
In (3+1)D, the $\chi_{\vek{p}}^\pm$ are four-component spinors
that encode the energy-band index and the Dirac particle's intrinsic
angular-momentum (spin) vector, where the corresponding spin
operator is defined as
\begin{equation}\label{eq:3+1Dspin}
\vek{S} = \frac{1}{4 i} \, \vek{\alpha} \times\vek{\alpha}
\equiv \frac{1}{2 i} \, \left( \begin{array}{c} 
    \alpha_2 \, \alpha_3 \\ 
    \alpha_3 \, \alpha_1 \\ 
    \alpha_1 \, \alpha_2  \end{array}
\right) \quad .
\end{equation}

\subsection{Symmetries}
The symmetries of the Dirac equation include operations that reflect
the properties of Minkowski space time. (Point group operations give
rise to the Lorentz group, whereas the Poincar\'e group also
includes translations in space and time.) These symmetries are also
relevant for a classical (non-quantum) description of relativistic
systems. On the other hand, in general we also have symmetries that
are present only in a quantum mechanical description of relativistic
systems. Each of these categories include discrete and continuous
symmetries.

In the following, we will focus mostly on the discrete symmetries
that are specific for the quantum-mechanical Dirac equation. Among
the spatial symmetries, we will concentrate on the nontrivial
operations of parity (space inversion) and rotations by $2\pi$.
Moreover, we will consider the operation of time reversal (also
called \emph{reversal of the motion\/}). \cite{sac87} The Hilbert
space of the (single-particle) Dirac equation includes solutions
with positive and negative energies corresponding to particles and
holes (i.e., anti-particles). These are related by
\emph{particle-hole conjugation\/} and \emph{energy-reflection
 symmetry\/}~\cite{gre00, zir96, alt97, ber02, cha08}. Lorentz
invariance will not feature at all.

We continue this Section with a brief introduction to particular discrete
symmetries and review their representation for the (3+1)D Dirac case.

\subsubsection{Spatial Symmetries --- in particular, Parity}
A spatial symmetry transformation $g$ is represented by a unitary
operator $\hat{P}(g)$ acting in the Hilbert space of the Hamiltonian
$H$, where the set $\{g\}$ of symmetry transformations forms a group
$G$. To describe the symmetry of an observable $\mathcal{O}$ under
the symmetry transformation $g$ we use the short-hand notation $g \,
\mathcal{O} \equiv \hat{P}(g) \, \mathcal{O} \, \hat{P}^{-1} (g)$.
When $\hat{P}(g)$ acts on the basis functions $\{ \ket{\nu}\}$ of
$H$, we can expand $\hat{P}(g) \ket{\nu}$ in terms of the same set of
basis functions $\{ \ket{\mu}\}$,
\begin{equation}
  \hat{P}(g) \, \ket{\nu} = \sum_\mu \ir (g)_{\mu\nu} \ket{\mu} \, ,
\end{equation}
so that the set of matrices $\{ \ir (g) \}$ forms a representation
for $G$. Then the invariance of the Hamiltonian $H$ under the
symmetry transformation $g$ implies~\cite{bir74}
\begin{equation}
  \label{eq:spatial-invar}
  \ir(g) \, H(\pp) \, \ir^{-1}(g)
  = H (g \, \pp) .
\end{equation}
This equation expresses the fact that we can view the effect of $g$
from two equivalent perspectives: we may regard it as a unitary
transformation of the basis functions $\{ \ket{\nu}\}$ of $H$
(``active'' view). This is equivalent to applying $g$ to momentum
$\pp$ (and position $\rr$, ``passive'' view). If $g$ is a symmetry
of $H$, both views must yield the same result. The invariance
condition (\ref{eq:spatial-invar}) holds generally for continuous
symmetry transformations (such as rotations) as well as discrete
operations (e.g., parity).

Equation (\ref{eq:spatial-invar}) includes as a special case
\emph{parity} $\mathcal{P}$ (i.e., space inversion). In $d=3$
spatial dimensions, it is characterized via the relations
\begin{subequations}
  \begin{gather}
    \mathcal{P}\,\rr = - \rr, \hspace{2em}
    \mathcal{P}\,\pp = - \pp \\[1ex]
    \ir(\mathcal{P}) \, H(\pp) \, \ir^{-1} (\mathcal{P})
    = H (\mathcal{P} \, \pp)
    = H (- \pp)
  \end{gather}
\end{subequations}
Note that, like $\rr$ and $\pp$, also the velocity $\vek{v}$ is odd
under parity. For the Dirac Hamiltonian in (3+1)D, the representation
matrix for the parity operation is
\begin{equation}\label{eq:Parity3+1D}
  \ir(\mathcal{P}) = \beta .
\end{equation}
As to be expected, we have $\ir (\mathcal{P}^2) = + \openone$.

\subsubsection{Rotations $\bm{\mathcal{R}}$}
In (3+1)D, the Hamiltonian $H_\mathrm{D}$ commutes with the operator
of total angular momentum $\vek{J} = \vek{L} + \vek{S}$, where
$\vek{L} = \vek{r} \times \vek{p}$ is the operator of orbital
angular momentum, and the spin operator $\vek{S}$ is given by Eq.\
(\ref{eq:3+1Dspin}).
Total angular momentum $\vek{J}$ is the generator for finite
rotations. Rotations $\mathcal{R}_{\hat{\vek{n}}}$ by $2\pi$ about
an arbitrary axis $\hat{\vek{n}}$ map the system onto itself but, in the
process, quantum states acquire a minus sign:
\begin{equation}
  \label{eq:rot2pi_3D}
  \ir (\mathcal{R}_{\hat{\vek{n}}} (2\pi))
  = \exp (2\pi i \,\hat{\vek{n}} \cdot \vek{J}) = - \openone \, .
\end{equation}
In that sense, the symmetry group becomes a double group which
is common for half-integer spin systems~\cite{wig59}.

\subsubsection{Time Reversal}
Time reversal (TR) is an anti-unitary transformation and can be
represented by the operator
\begin{equation}\label{eq:TRop}
  \theta = \mathcal{T} \mathcal{K}
\end{equation}
where $\mathcal{K}$ denotes complex conjugation, and the operator
$\mathcal{T}$ is the unitary transformation that relates the
time-reversed (i.e., complex-conjugated) basis spinors to the
original basis~\cite{bir74, win10a}. TR invariance requires that the
Dirac equation has the same solution as the equation obtained by
applying $\theta$ to both the basis functions and all tensors that
represent physical quantities (such as momentum $\pp$), giving
\begin{equation}
  \label{eq:time-invar}
  \ir (\mathcal{T}) \, H^\ast (\pp) \, \ir (\mathcal{T})^{-1}
  = H (\theta \,\pp)
  = H (-\pp) .
\end{equation}
Here, $\ast$ denotes complex conjugation. The minus sign reflects
the fact that momentum $\pp$ is odd under reversal of motion. This
implies that, as to be expected, $\theta$ reverts the velocity
$\vek{v}\equiv\vek{\alpha}$, $\ir (\mathcal{T}) \, \vek{\alpha}^\ast \,
\ir (\mathcal{T})^{-1} = -\vek{\alpha}$, but does not change the mass
term, $\ir (\mathcal{T}) \, \beta^\ast \, \ir (\mathcal{T})^{-1} =\beta$.
For the Dirac Hamiltonian in (3+1)D in the standard
representation~\cite{gre00} where only $v_2$ is imaginary, we have
\begin{equation}\label{eq:3DTR}
  \ir (\mathcal{T}) = -i\, v_1\, v_3 \equiv i\, \gamma^1\, \gamma^3 \quad ,
\end{equation}
which implies $\ir (\theta^2) = - \openone$. The remarkable property that
the representation matrix for the time reversal operation squares to the
\emph{negative\/} of the unit matrix for a (3+1)D Dirac particle arises as a
consequence of its intrinsic half-integer spin~\cite{sak94}.

\subsubsection{Particle-Hole Conjugation}
Particle-hole conjugation is also an anti-unitary
operation~\cite{gre00, zir96, alt97, ber02}. We can express it using
the operator
\begin{equation}\label{eq:phConj}
\kappa = \mathcal{C} \mathcal{K} \quad ,
\end{equation}
with a unitary operator $\mathcal{C}$. Particle-hole symmetry of a
Hamiltonian holds if the relation
\begin{equation}\label{eq:phConj-inv}
\ir (\mathcal{C}) \, H^\ast (-\pp) \, \ir(\mathcal{C})^{-1}
  = -H (\pp)
\end{equation}
is satisfied. The operation of particle-hole conjugation reverses
momentum and spin but leaves position invariant~\cite{gre00}. For
the Dirac Hamiltonian (\ref{eq:dirac_ham}) in $\pp$ representation, the
relation (\ref{eq:phConj-inv}) is equivalent to the conditions $\ir
(\mathcal{C}) \,\vek{\alpha}^\ast \, \ir(\mathcal{C})^{-1} = \vek{\alpha}$
and $\ir(\mathcal{C}) \, \beta^\ast\, \ir(\mathcal{C})^{-1} = -\beta$.
For the Dirac Hamiltonian in (3+1)D, we have~\cite{gre00}
\begin{equation}
  \ir (\mathcal{C}) = i \, \beta\, v_2 \equiv i \gamma^2 \quad  ,
\end{equation}
which implies $\ir (\kappa^2) = + \openone$.

\subsubsection{Chirality and Energy Reflection Symmetry}
In (3+1)D, an additional matrix $\gamma^5 = i \gamma^0\gamma^1
\gamma^2\gamma^3$ can be defined that anticommutes with all other
$\gamma^\mu$. In the massless limit $m=0$, $\gamma^5$ commutes with
the Dirac Hamiltonian
\begin{equation}
  \label{eq:chir_rel}
  [ H_\mathrm{D} (\pp,m=0), \gamma^5 ] = 0,
\end{equation}
so that it becomes the generator $\chi$ for a continuous chiral
symmetry~\cite{gre00} with $\ir (\chi) = \gamma^5$.

In the context of random-matrix theory (RMT), the term \emph{chiral
 symmetry} has departed from its original meaning in relativistic
quantum mechanics~\cite{alt97,ber02}. The chirality matrix
$\gamma^5$ anticommutes also with the \emph{Dirac operator}
$D = \gamma^\mu \, P_\mu - m$, see Eq.~(\ref{eq:dirac_op}).
Except for zero modes, the eigenfunctions of $D$ occur in pairs
$u_n$, $\gamma^5 u_n$ with opposite eigenvalues ($\lambda_n,
-\lambda_n$). This notion \cite{leu92} has been transfered to other
systems. In RMT, a \emph{chiral} symmetry~\cite{alt97, ber02} of
a Hamiltonian $H$ is associated with a unitary operator $\mathcal{M}$
that satisfies
\begin{equation}
  \label{eq:chiral-sign}
  \ir (\mathcal{M}) \, H(\pp) \, \ir^{-1}
  (\mathcal{M}) = - H(\pp) \quad .
\end{equation}
This operation reverses the energy. In recent studies~\cite{cha08},
the operator $\mathcal{M}$ has therefore been called
\emph{energy-reflection} (ER) symmetry. In order to distinguish
$\mathcal{M}$ from the chiral symmetry (\ref{eq:chir_rel}), we will
follow the latter naming convention. The operator $\mathcal{M}$
reverses the velocity, while leaving momentum and position invariant:
$\mathcal{M} \, \pp = \pp$ and $\mathcal{M} \, \vek{r} = \vek{r}$. For
the Dirac Hamiltonian, the corresponding representation matrix $\ir
(\mathcal{M})$ must therefore anticommute with the matrices
$\vek{\alpha}$ and $\beta$.

For the (unperturbed) Dirac equation (\ref{eq:dirac}), time
reversal symmetry (\ref{eq:time-invar}), particle-hole conjugation
(\ref{eq:phConj-inv}) and the energy reflection symmetry
(\ref{eq:chiral-sign}) are not independent symmetries~\cite{ber02}.
We have
\begin{equation}
  \mathcal{M} = \kappa \, \theta
\end{equation}
giving
\begin{equation}\label{eq:ERrelate}
  \ir(\mathcal{M}) = \ir(\mathcal{C})\, \left[ \ir(\mathcal{T})\right]^\ast
  \quad ,
\end{equation}
and explicitly in (3+1)D
\begin{equation}\label{eq:ERchi}
  \ir(\mathcal{M}) = - i \, \gamma^0\, \gamma^5
  \equiv - i \, \beta \, \ir (\chi) \quad ,
\end{equation}
which implies $\ir (\mathcal{M}^2) = + \openone$.

\begin{figure}[b]
\centering
\includegraphics[width=0.75\columnwidth]{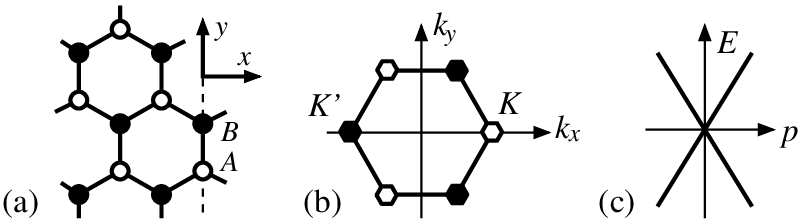}
\caption{\label{fig:graphene}Basic structure and electronic
properties of graphene. (a)~Schematics of a honeycomb lattice
with sites on the two sublattices $A$ and $B$ indicated by empty
and filled circles, respectively. (b)~First Brillouin zone of the
honeycomb lattice, with high-symmetry points $K$ and $K'$ indicated.
(c)~The electron energy $E$ varies linearly as a function of
crystal momentum $p$ near a $K$ or $K'$ point, mimicking
the behaviour of massless Dirac particles.}
\end{figure}

\subsection{Physical motivation: Electronic properties of graphene}
The dynamics of electrons in crystalline solids is dictated by the
material's band structure~\cite{ros09}, i.e., the energy of the
electrons as a function of crystal momentum (or quasi-momentum)
$\vek{p}$. Near high-symmetry
points in the Brillouin zone, the energy bands are often parabolic,
defining an effective mass for the quasi-free crystal electrons that behave
qualitatively like electrons in vacuum~\cite{ros09}. However, in some
materials, the band structure deviates markedly from this usual
situation. A single sheet of graphite, called graphene, is such an
example~\cite{gei09}. It consists of carbon atoms arranged on a
honeycomb lattice as illustrated in Fig.~\ref{fig:graphene}(a).
Inspection shows that such a lattice has two equivalent sublattices,
labelled $A$ and $B$, which are represented by a pseudospin-1/2
degree of freedom~\cite{cas09}. Near a $K$ point in the Brillouin
zone [Fig.~\ref{fig:graphene}(b)], the energy dispersion turns out to
be linear and the electronic eigenstates are chiral in pseudo-spin
space, similar to ultra-relativistic (i.e., massless) Dirac particles.
As the same holds true near the non-equivalent $K'$ point, the
low-energy description of electrons in graphene is a realisation of
a two-flavour (2+1)D Dirac model~\cite{jac81, sem84, app86, dit00,
gus07}, where association with a $K$ or $K'$ point defines the two
flavours.

\section{(2+1)-dimensional Dirac theory}
In this Section, issues relating to the discrete symmetries for the
(2+1)D Dirac theory are discussed. In (2+1)D spacetime, the Dirac
Hamiltonian becomes a $2 \times 2$ matrix. In this case, we have
only four linearly independent basis matrices that we may choose as
Pauli matrices~\cite{sak94} $\sigma_x$, $\sigma_y$, and $\sigma_z$,
as well as $\sigma_0 \equiv \openone_{2 \times 2}$. Yet we still
need to represent intrinsic angular momentum (spin) and two energy
bands. The reduced number of basis matrices thus gives rise to
pathologies that prevent the consistent definition of many
symmetries, except in the mass-less limit ($m=0$).

Explicitly, we can choose a representation where
\begin{equation}\label{eq:2D_Dirac}
\vek{\alpha} = (\sigma_x, \sigma_y) \quad \mbox{and}\quad
\beta = \sigma_z \, .
\end{equation}
Our representation is equivalent, but not identical, to previously
discussed ones~\cite{jac81, sem84, app86}, as a look at the $\gamma$
matrices shows:
\begin{equation}\label{eq:2D_Dirac_gamma}
\gamma^0 = \sigma_z \,\, , \quad \gamma^1 = i\sigma_y\,\, , \quad
\gamma^2 = -i\sigma_x \,\,  .
\end{equation}

\subsection{Angular Momentum and Rotations $\bm{\mathcal{R}}$}
Straightforward calculation establishes that the (2+1)D version of
$H_\mathrm{D}$ commutes with the operator of total angular momentum
$J_z = L_z + S_z$, where $L_z = x\, p_y - y\, p_x$ is the $z$
component of orbital angular momentum, and
\begin{equation}\label{eq:spin}
S_z \equiv \frac{1}{4 i} \, \left(\vek{\alpha} \times\vek{\alpha} \right)
 \cdot \hat{\vek{z}}  = \frac{1}{2}\, \sigma_z\quad .
\end{equation}
Thus the intrinsic spin degree of freedom formally emerges in
complete analogy to the (3+1)D case; see Eq.~(\ref{eq:3+1Dspin}) and
Ref.~\cite{tha92}, but only its $z$ component is a relevant
quantity in (2+1)D.

Rotations $\mathcal{R}_z$ by $2\pi$ about the $z$ axis map the system
onto itself with a minus sign
\begin{equation}
  \label{eq:rot2pi_2D}
  \ir (\mathcal{R}_z (2\pi)) = \exp (2\pi i J_z) = - \openone \, ,
\end{equation}
similar to Eq.\ (\ref{eq:rot2pi_3D}), so that once again
the symmetry group becomes a double group~\cite{wig59}.

\subsection{Parity $\bm{\mathcal{P}}$}
In three spatial dimensions, parity $\mathcal{P}$ is defined as
space inversion $\rr \rightarrow -\rr$. Its interesting aspects
arise from the fact that the determinant of the associated
representation matrix is $\det (\ir(\mathcal{P})) = -1$, so that
parity cannot be expressed in terms of infinitesimal unitary
transformations. In two spatial dimensions, the transformation $\rr=
(x,y) \rightarrow -\rr=(-x,-y)$ is not interesting because it is
equivalent to a rotation by $\pi$ and the determinant of the
representation matrix is $+1$. This is why parity in 2D is
usually defined~\cite{jac81,app86,dit00,sem84} as either
\begin{subequations}
  \label{eq:parity}
  \begin{equation}
    \label{eq:parity_x}
    \rr=(x,y) \rightarrow \mathcal{P}_x \, \rr = (x,-y)
  \end{equation}
  or
  \begin{equation}
    \label{eq:parity_y}
    \rr=(x,y) \rightarrow \mathcal{P}_y \, \rr = (-x,y) \quad .
  \end{equation}
\end{subequations}
The same transformation laws apply for linear momentum $\pp$ and
velocity $\vek{v}$. Clearly, parity invariance as a property of a
physical operator should be established with respect to both
operations (\ref{eq:parity_x}) and (\ref{eq:parity_y}) to count as a
proper symmetry. The above 2D versions of parity imply that $L_z$ is
odd under parity,
\begin{equation}
  \label{eq:parity-orb-ang}
  \mathcal{P}_x \, L_z = \mathcal{P}_y \, L_z = - L_z \quad .
\end{equation}

If we require that $\vek{\alpha}\cdot\pp$ is even under parity, we find
\begin{equation}
  \ir(\mathcal{P}_\nu) = \sigma_\nu \qquad \mbox{with} \qquad
  \nu = x,y \quad .
\end{equation}
Furthermore, this implies~\cite{jac81} that $S_z$ and $J_z$ are
likewise odd under parity. In contrast, $\beta$ must be even under
parity. As $S_z = \beta/2$ for the (2+1)D Dirac theory, it is
impossible to satisfy both conditions, i.e., the mass term violates
parity symmetry in (2+1)D. Only in the mass-less limit $m=0$,
$H_\mathrm{D}$ is invariant under parity.

We can study parity also for the $\psi_\pp^\pm (\rr)$ given in
Eq.~(\ref{eq:eigenstates}), which are eigenfunctions for both the
Dirac Hamiltonian and linear momentum $\vek{p} = (p_x, p_y)$. (See
also Problem 3.4 in Ref.~\cite{sak67}.) In the $m=0$ limit and for
$E^2 = p_x^2 + p_y^2 \ne 0$, we have (apart from a normalization
constant)
\begin{equation}
  \label{eq:free_wf2}
  \psi_\pp^\pm (\rr) \propto 
  \ee^{i \pp \cdot \rr} \tvek{E_\pm \\ p_x+ip_y}
\end{equation}
If parity is defined as above it follows that
\begin{equation}
  \label{eq:free_wf2_parity}
  \ir(\mathcal{P}_\nu) \, \psi_\pp^\pm (\rr) =
  \eta_\pp^\pm \,
  \psi_{\mathcal{P}_\nu^{-1} \pp}^\pm (\mathcal{P}_\nu^{-1} \, \rr) \,\, .
\end{equation}
with a phase factor $\eta_\pp^{\pm}$. (Of course, we have
$\mathcal{P}_\nu^{-1} = \mathcal{P}_\nu$.)
Similar to Eq.\ (\ref{eq:spatial-invar}), this equation illustrates
the fact that we can view a symmetry transformation such as parity
from two equivalent perspectives: we may regard it as a unitary
transformation of the expansion coefficients (i.e., the spinors).
This is equivalent to the inverse coordinate transformation, where
position $\rr$ and momentum $\pp$ are mapped on the inversely
transformed quantities. As expected, parity flips position vectors
and momenta, while the energy is preserved. Also, the expectation
value of $S_z$ is reversed under parity.

\subsection{Time reversal $\bm{\theta}$}
The basic requirements for TR invariance imply that $\beta$ should
be even under TR while $S_z$ should be odd. As in the case of
parity, the fact that both quantities are represented by $\sigma_z$
in the (2+1)D situation makes it impossible to find such a TR
transformation, i.e., the mass term violates both parity and TR
symmetry. However, TR can be defined consistently in the massless
limit: the usual~\cite{sak94}
\begin{equation}
  \label{eq:timrev}
\theta = \mathcal{T} \mathcal{K}
\hspace{1em} \mbox{with} \hspace{1em}
\ir (\mathcal{T}) = i \sigma_y \quad ,
\end{equation}
where $\sigma_y$ is the imaginary Pauli matrix, implies
\begin{equation}
  \theta \tvek{a_1 \\ a_2} = \ir (\mathcal{T}) \tvek{a_1^\ast \\ a_2^\ast}
  = \tvek{a_2^\ast \\ - a_1^\ast}
\end{equation}
for arbitrary spinors $\tvek{a_1 \\ a_2}$ and $\ir (\theta^2) = -
\openone$. More explicitly, we obtain for the wave function
(\ref{eq:free_wf2})
\begin{equation}
  \label{eq:timerev-wf}
  \theta \, \psi_\pp^\pm
  = \ir (\mathcal{T}) \, \psi_\pp^{\pm\,\ast}
  = \zeta_\pp^\pm \, \psi_{-\pp}^\pm
\end{equation}
with a phase factor $\zeta_\pp^\pm$,
i.e., momentum is reversed, while the energy is preserved. 
Also, the expectation value of $S_z$ is reversed under TR.

\subsection{Particle-hole conjugation $\bm{\kappa}$}
As $\alpha_x$ and $\beta$ are real matrices in our adopted representation
of the (2+1)D Dirac theory while $\alpha_y$ is imaginary, the matrix
$\mathcal{C}$ entering the definition (\ref{eq:phConj}) of
particle-hole conjugation needs to commute with $\alpha_x$ and
anticommute with both $\alpha_y$ and $\beta$. These conditions are
satisfied by the matrix $\sigma_x$. Hence, the
particle-hole-conjugation operation $\kappa = \mathcal{C}
\mathcal{K}$ can be consistently defined for the (2+1)D case via the
representation matrix
\begin{equation}
\ir (\mathcal{C}) =\sigma_x \quad 
\end{equation}
so that $\ir (\kappa^2) = +\openone$.

\subsection{Energy Reflection Symmetry $\bm{\mathcal{M}}$}
ER symmetry requires $\ir(\mathcal{M})$ to anticommute with all
Dirac matrices (i.e., the components of $\vek{\alpha}$ and $\beta$). No
such matrix exists in (2+1)D and, hence, ER symmetry cannot be
established for the Dirac model. However, $\ir(\mathcal{M})=
\sigma_z$ satisfies the ER condition (\ref{eq:chiral-sign}) in the
mass-less limit with $\ir (\mathcal{M}^2) = +\openone$.

\subsection{Chiral symmetry $\bm{\chi}$}
The fact that no matrix anticommutes with all three Pauli matrices,
i.e., the $\gamma^\mu$ given in Eq.~(\ref{eq:2D_Dirac_gamma}), also
implies that no equivalent of $\gamma^5$ exists in
(2+1)D~\cite{jac81, app86}. As a result, no chiral symmetry $\chi$
can be established, \emph{even in the mass-less limit\/}.

\subsection{Discussion}
It appears that particle-hole conjugation and spatial rotations are
the only symmetries that can reasonably be considered within the
(2+1)D Dirac theory. The fact that consistent parity, TR and ER
operations exist only in the zero-mass limit, and that chiral
symmetry is altogether impossible to define, makes this theory
highly pathological from a physical point of view. As it turns out,
this unsatisfactory situation can be remedied by introducing an
additional flavour degree of freedom for the Dirac
fermions~\cite{jac81}. This generalization of (2+1)D Dirac theory
will be explored in the next Section.

\section{Two-flavour (2+1)D Dirac theory}\label{sec:2f3d}
It is possible to construct a theory of Dirac fermions in (2+1)D
that exhibits the familiar symmetries of parity and TR, even for a
finite mass~\cite{jac81}. This theory involves four-spinor wave
functions describing two flavours (2Fs) of such fermions. The Dirac
operator (\ref{eq:dirac_ham}) is block-diagonal, and a possible choice
for the matrices $\alpha_j$ and $\beta$ is
\begin{equation}
  \vek{\alpha} = \left( \! \tvek[cc]{ \sigma_x & 0 \\ 0 & \sigma_x},
                   \tvek[cc]{ \sigma_y & 0 \\ 0 & \sigma_y} \!\right)
  \hspace{1em} \mbox{and} \hspace{1em}
  \beta = \tvek[cc]{ \sigma_z & 0 \\ 0 & - \sigma_z} .
\end{equation}
Such a flavour-symmetric representation was adopted, e.g., in
Refs.~\cite{ale06, akh07, bee08, wur12} to describe the
electronic degrees of freedom near the two valleys in graphene.
Other formulations found in the literature~\cite{jac81, sem84,
 app86, dit00, gus07, cha08, wur09} are unitarily equivalent to the one
used here. According to Eq.\ (\ref{eq:spin}), the spin operator
$S_z$ emerges as
\begin{equation}
S_z = \frac{1}{2} \tvek[cc]{ \sigma_z & 0 \\ 0 & \sigma_z } \quad .
\end{equation}
Thus, in contrast to the single-flavour case, $\beta$ and $S_z$ are
\emph{not\/} proportional to each other in the 2F (2+1)D Dirac
theory. For completeness, we also give the $\gamma$ matrices:
\begin{equation}
\gamma^0 = \beta,  \quad
\gamma^1 = \tvek[cc]{ i \sigma_y & 0 \\ 0 & - i \sigma_y}, \quad
\gamma^2 = \tvek[cc]{ -i \sigma_x & 0 \\ 0 &  i \sigma_x} \, .
\end{equation}

\subsection{Symmetries}
Rotations are again generated by $J_z = L_z + S_z$, which is a
conserved quantity. Hence, as in the single-flavour case, a $2\pi$
rotation is equivalent to $-\openone$, which is a signature of
half-integer spin.

We define parity, TR and ER as symmetry transformations that connect
the flavour subspaces. For each of the parity transformations
$\mathcal{P}_{x,y}$, two inequivalent representation matrices $\ir_\pm
(\mathcal{P}_\nu)$ can be realised,
\begin{equation}
  \ir_\pm (\mathcal{P}_\nu)
  = \tvek[cc]{0 & \sigma_\nu \\ \pm\sigma_\nu & 0} \; ,
\end{equation}
which are distinguished by the property $\ir_\pm (\mathcal{P}_\nu^2)
=\pm\openone$. Similarly, the anti-unitary operator $\theta =
\mathcal{T}\, \mathcal{K}$ with representation matrices
\begin{equation}
  \label{eq:timrev-flav}
  \ir_\pm (\mathcal{T})
  = \tvek[cc]{ 0 & -i \sigma_y \\ \pm i  \sigma_y & 0}
\end{equation}
keeps the 2F (2+1)D Dirac Hamiltonian invariant while reversing
velocity, momentum and spin. Action on a general state yields
\begin{equation}
  \label{eq:timrev-flav-genvec}
  \ir_\pm (\theta) \tvek{a_1 \\ a_2 \\ a_3 \\ a_4}
  = \tvek{- a_4^\ast \\ a_3^\ast \\ \pm a_2^\ast \\ \mp a_1^\ast} \quad ,
\end{equation}
which implies $\ir_\pm (\theta^2) =\pm\openone$. Thus it is
possible~\cite{win10} that Dirac particles in 2F (2+1)D behave under
TR like spinless particles [having $\ir (\theta^2) = +\openone$] as
an alternative to the expected spin-1/2-behaviour [where $\ir
(\theta^2) = -\openone$].

The conditions associated with invariance under particle-hole
conjugation $\kappa = \mathcal{C}\, \mathcal{K}$ are satisfied by
\begin{equation}
  \ir_\pm (\mathcal{C}) = \tvek[cc]{\sigma_x & 0 \\ 0 & \mp \sigma_x} \quad .
\end{equation}
Unlike the TR and chirality operations, particle-hole conjugation
preserves the flavour degree of freedom so that we get $\ir_\pm
(\kappa^2) = + \openone$.

Finally, there exist two inequivalent representation matrices for ER
symmetry
\begin{equation}
\ir_\pm (\mathcal{M}) = \tvek[cc]{ 0 & \sigma_z \\ \pm \sigma_z
& 0} \quad .
\end{equation}
They both anticommute with all Dirac matrices in the 2F (2+1)D
model, satisfy the relation $\ir_\varepsilon (\mathcal{M}) = \ir_\pm
(\mathcal{C}) \, \ir_{\pm\varepsilon} (\mathcal{T})^\ast$ [$\varepsilon
=\pm 1$; cf.\ also Eq.~(\ref{eq:ERrelate})] and, similarly to the
case of the other flavour-coupling discrete symmetry
transformations, are distinguishable by the sign of their squares:
$\ir_\pm (\mathcal{M}^2) =\pm\openone$.

A continuous chiral symmetry also exists in 2F (2+1)D
models~\cite{app86, cha08}. In our representation, the matrices
\begin{equation}
\ir_\pm (\chi) = \tvek[cc]{0 & \sigma_0 \\ \pm \sigma_0 & 0} 
\end{equation}
anticommute with $\beta$ and commute with $\alpha_{x,y}$, thus satisfying
the condition for the generator of a chiral symmetry. Again, the two
matrices are distinguishable by the sign of their squares: $\ir_\pm
(\chi^2) = \pm\openone$. It is possible to express the $\ir_\pm
(\chi)$ in terms of other Dirac operators and the ER
transformations,
\begin{equation}
\ir_\pm (\chi) = -i\, \alpha_x \, \alpha_y \, \ir_\pm
(\mathcal{M}) = \beta \, \ir_\mp(\mathcal{M}) \quad ,
\end{equation}
which mirrors Eq.~(\ref{eq:ERchi}).

\subsection{Discussion}
\begin{table}[t]
  \caption{\label{tab:2Fops}Summary of 2F (2+1)D operators given in
   compact notation where Pauli matrices $\sigma_j$ and $\tau_j$ act
   in Dirac and flavour space, respectively, and $\sigma_0 = \tau_0 =
   \openone_{2\times 2}$.}
\centering
\begin{tabular}{CCCCCCCCCCCC}
\bottomrule\toprule
      \alpha_x & \alpha_y & \beta\equiv \gamma^0 & \gamma^1 & \gamma^2 & 2 S_z
      \\ \sigma_x\otimes\tau_0 & \sigma_y\otimes\tau_0 & \sigma_z
      \otimes\tau_z & i\sigma_y\otimes\tau_z & -i \sigma_x\otimes\tau_z &
      \sigma_z \otimes\tau_0 & \\ \bottomrule \toprule
      \ir_+(\mathcal{P}_\nu) & \ir_-(\mathcal{P}_\nu) & \ir_+ (\mathcal{T}) &
      \ir_- (\mathcal{T}) & \ir_+(\mathcal{C}) & \ir_-(\mathcal{C}) & \\
      \sigma_\nu\otimes\tau_x & i \sigma_\nu\otimes\tau_y & \sigma_y
      \otimes\tau_y & -i \sigma_y\otimes \tau_x & \sigma_x\otimes\tau_z &
      \sigma_x\otimes\tau_0 \\ \bottomrule\toprule
      \ir_+(\mathcal{M}) & \ir_-(\mathcal{M}) & \ir_+ (\mathcal{\chi}) &
      \ir_- (\mathcal{\chi}) \\ \sigma_z\otimes\tau_x & i \sigma_z
      \otimes\tau_y & \sigma_0\otimes\tau_x & i \sigma_0\otimes\tau_y
      \\ \bottomrule\toprule
   \end{tabular}
\end{table}
For reference and to facilitate easier comparison with the
literature, we provide expressions of relevant operators for the 2F
(2+1)D Dirac model using a compact notation (with Pauli matrices
$\sigma_j$ and $\tau_j$ acting in Dirac and flavour space,
respectively) in Table~\ref{tab:2Fops}.

The existence of two possible realisations for each of the
symmetries $\mathcal{S} = \mathcal{P}_\nu, \mathcal{T}, \mathcal{C},
\mathcal{M}$, and $\chi$ seems puzzling, especially for many of
those transformations where the squares of the two realisations have
opposite signs. However, for any specific physical realisation of a
2F (2+1)D Dirac system, the transformation properties of the basis
functions will uniquely determine for each $\mathcal{S}$ which of
the two operators $\ir_\pm(\mathcal{S})$ corresponds to the actual
symmetry in this system. For fermionic (spin-1/2) particles, TR will
be represented by~\cite{sak94} $\ir_- (\mathcal{T})$. In contrast,
$\ir_+ (\mathcal{T})$ applies to electrons in graphene~\cite{bee08}
whose low-energy band structure (neglecting the real spin) realises
a 2F (2+1)D Dirac model, with the valley index being the flavour
degree of freedom~\cite{cas09}.

In previous discussions~\cite{app86} of chiral symmetry in 2F (2+1)D
Dirac systems, the existence of two possible representations
$\ir_\pm(\chi)$ was seen to imply the existence of a U(2) symmetry
with generators $\{\openone, \ir_+(\chi), \ir_-(\chi),
\ir_+(\chi)\ir_-(\chi) \}$. Considering also the other pairs of
representation matrices $\ir_\pm (\mathcal{S})$, we find more
generally
\begin{equation}
\ir_+ (\mathcal{S}) \, \ir_- (\mathcal{S}) =
\tvek[cc]{ -\openone & 0 \\ 0 & \openone} \, .
\end{equation}
for each symmetry $\mathcal{S}$. Hence the product of the two
possible representation matrices for each symmetry acts differently
in the subspaces associated with the two flavours. Therefore, in
situations where the two flavours are physically indistinguishable
(which is the case, e.g., in an ideal sheet of graphene), the two
representation matrices $\ir_\pm (\mathcal{S})$ cannot be
interpreted as distinct discrete symmetries of this system.

\section{(1+1)-dimensional Dirac theory}
For comparison, we briefly discuss the (1+1)D Dirac model where $\pp
\equiv p$ in $H_\mathrm{D}$ given by Eq.~(\ref{eq:dirac_ham}). For
definiteness we choose $\vek{\alpha} = \sigma_x$ and $\beta = \sigma_z$.
Parity requires
\begin{equation}
   \ir (\mathcal{P}) = \sigma_z
\end{equation}
with $\ir (\mathcal{P}^2) = + \openone$. For TR, we find $\theta =
\mathcal{T}\, \mathcal{K}$ with
\begin{equation}
  \ir (\mathcal{T}) = \sigma_z \quad .
\end{equation}
Thus we have $\ir (\theta^2) = + \openone$, consistent with the fact
that a Dirac particle in (1+1)D is a spinless object~\cite{tha92}.
Particle-hole conjugation is realised by $\kappa = \mathcal{C}
\,\mathcal{K}$ where
\begin{equation}
  \ir (\mathcal{C}) = \sigma_x
\end{equation}
with $\ir (\mathcal{\kappa}^2) = + \openone$. The ER transformation with
\begin{equation}
  \ir(\mathcal{M}) = -i \, \sigma_y
\end{equation}
and $\ir (\mathcal{M}^2) = + \openone$ satisfies the condition
(\ref{eq:chiral-sign}), and $\ir (\chi) = \sigma_x$ defines a chiral
symmetry.

\section{Conclusions \& Outlook}
We investigated basic properties of discrete symmetry operations for
low-dimensional Dirac models. An additional flavour degree of
freedom is needed to introduce the usual discrete symmetries known
from Dirac particles in three spatial dimensions. For the
two-flavour (2+1)-dimensional Dirac model two inequivalent
representation matrices exist for each symmetry operation.
With the exception of particle-hole conjugation, the two
representation matrices for each symmetry are distinguishable by
the sign of their squares.
 
Symmetry analyses such as that presented in this work enable the
classification of perturbations to the (2+1)-dimensional Dirac
model~\cite{ber02}, making it possible to draw conclusions about how
the perturbations affect the spectrum and, ultimately, the physical
properties of condensed-matter realisations such as graphene
sheets~\cite{gus07}. The origin of such perturbations could be
disorder~\cite{ale06, mcc06a, ost06} or $\pp$-dependent corrections
to the Hamiltonian reflecting the lower crystal symmetry of, e.g., the
graphene lattice~\cite{win10a}.

\acks   This work was supported by Marsden Fund contract no.\ VUW0719,
  administered by the Royal Society of New Zealand. Useful discussions
  with M.~Visser and K.~Ziegler are gratefully acknowledged. U.~Z.\
  thanks the Departement Physik at the University of Basel, Switzerland,
  for hospitality during the writing of this article.


\end{document}